\documentclass[epj]{svjour}
\usepackage{graphicx}
\usepackage{bbm}
\usepackage{amsmath,amssymb}
\allowdisplaybreaks
\usepackage{bbm}
\usepackage{slashed}
\usepackage{pdflscape}
\usepackage{hyperref}

\usepackage[T1]{fontenc} 

\usepackage{dcolumn}   
\usepackage{bm}        
\usepackage{bbold}
\usepackage{epsfig}
\usepackage{multirow}
\usepackage{color}
\usepackage[normalem]{ulem}
\usepackage{multirow}

\newcommand{\gsim}{\raise.3ex\hbox{$>$\kern-.75em\lower1ex\hbox{$\sim$}}}
\newcommand{\lsim}{\raise.3ex\hbox{$<$\kern-.75em\lower1ex\hbox{$\sim$}}}

\newcommand{\nn}{\nonumber}

\newcommand{\half}{\frac{1}{2}}

\begin{document}
	%
	\title{Revealing timid pseudo-scalars with taus at the LHC}
	\author{Giacomo~Cacciapaglia\inst{1} \and Gabriele Ferretti\inst{2} \and Thomas Flacke\inst{3} \and Hugo Ser\^{o}dio\inst{4}
	}                     
	%
	%
	\institute{Universit\'{e} de Lyon, France; Universit\'{e} Lyon 1, CNRS/IN2P3, UMR5822, IPNL F-69622 Villeurbanne Cedex, France
		\and  Department of Physics, Chalmers University of Technology, Fysikg\aa rden, 41296 G\"oteborg, Sweden
		\and  Center for Theoretical Physics of the Universe, Institute for Basic Science (IBS), Daejeon, 34126, Korea
		\and  Department of Astronomy and Theoretical Physics, Lund University, SE-223 62 Lund, Sweden}
	\date{}
	%
	\abstract{A light pseudo-scalar that is copiously produced at the LHC may still be allowed by present searches. While masses above $65$ GeV are effectively covered by di-photon searches, the lower mass window can be tested by a new search for boosted di-tau resonances. We test this strategy on a set of composite Higgs models with top partial compositeness, where most models can be probed with an integrated luminosity below $300$ fb$^{-1}$.
		\PACS{
			{}{}
		} 
	} 
	\maketitle
\flushbottom

\section{Introduction}

The search for new resonances is one of the main physics goals at the LHC, with the discovery of a Higgs boson being an illustrious example \cite{Aad:2012tfa,Chatrchyan:2012xdj}. The efforts continue, mainly focusing on high mass objects typically heavier than the Higgs itself. There are in fact few searches exploring invariant masses of two Standard Model (SM) particles below, say, 100 GeV: one notable case is the search for a di-photon resonance \cite{Aad:2014ioa,CMS-PAS-HIG-17-013}, mostly motivated by models that feature an extended Higgs sector, like two Higgs doublet models~\cite{Cacciapaglia:2016tlr} and the next-to-minimal supersymmetric SM~\cite{Ellwanger:2015uaz}.

In this article, we focus on the LHC phenomenology of a light new scalar with a mass between $10$ and $100$ GeV. Generically, light new scalars are strongly constrained from electroweak precision measurements (indirectly) and from direct searches at LEP and Tevatron. At the LHC, besides the above-mentioned di-photon channel, light (pseudo)-scalars are usually searched for in the decays of the 125 GeV Higgs boson. Below roughly 10 GeV, strong bounds arise from searches related to mesons, or in experiments looking for light axion-like particles (ALPs)~\cite{Brivio:2017ije,Bellazzini:2017neg,Bauer:2017ris,Mariotti:2017vtv}.  Thus, the common lore is that a new scalar, in order to escape detection, needs to be either very heavy or weakly coupled to the SM.

Note, however, that it is enough to have small couplings to electrons and to the electroweak gauge bosons in order to escape direct LEP searches and electroweak precision bounds, as well as small couplings to the Higgs to avoid the Higgs portal constraints. Couplings to gluons (and heavy quarks) are less constrained, and may lead to sizable production rates at the LHC. Candidates of this kind arise naturally in composite Higgs models that enjoy a fermion-gauge underlying description~\cite{Barnard:2013zea,Ferretti:2013kya,Ferretti:2016upr,DeGrand:2016pgq,Belyaev:2016ftv} providing a partial UV completion. Recent lattice results~\cite{Ayyar:2017qdf} have started to address the mass spectrum in a specific model~\cite{Ferretti:2014qta}.

In this article, we will consider this class of models to explore the 10 to 100 GeV mass window and show that it is, in fact, poorly tested. A {\it timid} composite pseudo-scalar (TCP) arises as the pseudo-Nambu-Goldstone boson associated with an anomaly-free U(1) global symmetry in all models of partial compositeness that enjoy a UV completion, as defined in Ref.~\cite{Ferretti:2013kya}. All the possible models can be classified, and give precise predictions for the properties of the TCP candidate~\cite{Belyaev:2016ftv}, thus mapping out a complete landscape of possibilities.
We show that, while some models are already partly tested by the low mass di-photon searches, others are unconstrained. We point out that searches for boosted di-tau resonances (which could reach a lower invariant mass than the current value of 90 GeV~\cite{CMS-PAS-HIG-16-006,CMS-PAS-HIG-16-037}) give very promising signals and could be a powerful complementary probe to the di-photon channel, or even be the only way to access this class of TCPs.

\section{Description of the models}

The effective Lagrangian we consider is the SM Lagrangian augmented by the following terms, up to dimension five operators (counting powers of $f_a$):
\begin{align}
 {\mathcal{L}} &= \half(\partial_\mu a)(\partial^\mu a) -\half m_a^2 a^2 - \sum_f \frac{i C_f m_f}{f_a} a \bar \Psi_f \gamma^5 \Psi_f \label{mother} \\
 &\!\!\!\!\!\!+\!\frac{g_s^2 K_g a}{16\pi^2 f_a}G^a_{\mu\nu}\tilde G^{a\mu\nu}\!+\! \frac{g^2 K_W a}{16\pi^2 f_a} W^i_{\mu\nu}\tilde W^{i\mu\nu}
 \!+\!\frac{g'^2 K_B a}{16\pi^2 f_a} B_{\mu\nu}\tilde B^{\mu\nu}\nn \, .
\end{align}
A pseudo-scalar $a$ described by this general Lagrangian arises, for example, in UV completions of composite Higgs models, which were classified and studied in Refs \cite{Ferretti:2013kya,Belyaev:2016ftv}. In this section, we briefly summarize the main results relevant for our phenomenological study. Further background information on the models is provided in Appendix \ref{app:models}.

Within this class of models, the coupling to the SM fermions in Eq.~(\ref{mother}) is only the first term of the expansion of the spurion coupling  $   - m_f(h)\ \mathrm{e}^{i C_f a/f_a} \bar\Psi_{fL} \Psi_{fR} + \mathrm{h.c.}$ (generating the fermions masses), which breaks \emph{explicitly} the $U(1)$ shift symmetry.  A derivative coupling of the TCP to fermions of the form $(\partial_\mu a / f_a)  \bar \Psi_f \gamma^5 \gamma^\mu \Psi_f$ is absent in these models since the SM fermions are neutral under the TCP $U(1)$ charge. Although such a coupling can be obtained by using the fermion equations of motion on the leading term given in Eq.~(\ref{mother}), the two couplings are of genuinely different origin \cite{Kim:1984pt}, as manifested in the higher-order expansion of the spurion coupling. Starting from the complete spurion term, couplings of the Higgs to two TCPs, as well as to one TCP and $Z$ boson, arise at loop level and are given by (see Appendix \ref{app:loops} for the derivation)
\begin{eqnarray}
 \mathcal{L}_{haa}  &=&  \frac{3 C_t^2 m_t^2 \kappa_t}{8 \pi^2 f_a^2 v} \log \frac{\Lambda^2}{m^2_t}\ h (\partial_\mu a) (\partial^\mu a),\label{eq:haa}\\
 \mathcal{L}_{hZa} &=&  \frac{3 C_t m_t^2 g_A}{2 \pi^2 f_a v} (\kappa_t - \kappa_V) \log \frac{\Lambda^2}{m^2_t}\ h (\partial_\mu a) Z^\mu, \label{eq:hZa}
 \end{eqnarray}
where we list only the effect of the log-divergence ($\Lambda \sim 4 \pi f_a$), $g_A= -g/(4 \cos \theta_W)$ is the axial coupling of the $Z$ to tops, and $\kappa_{V,t}$ are the corrections from compositeness to the couplings of the Higgs to vectors and tops, respectively.
As $\kappa_{V,t} = 1 + \mathcal{O} (v^2/f_a^2)$, our result agrees with the fact that the only non-zero contribution to the $hZa$ coupling arises from a dimension 7 operator~\cite{Bauer:2016zfj}.

The couplings to gauge bosons in Eq.~(\ref{mother}) arise as anomalous couplings if the TCP is a (SM singlet) bound state of underlying SM charged fermions. In this case, the anomaly coefficients $K_{g,W,B}$ are fully determined by the charges of the hyper-fermions. We refer to \cite{Belyaev:2016ftv} for an extensive description of a classification of UV completions giving rise to this TCP, which yields twelve models. For the purpose of this article, the TCP dynamics in the twelve models is fully specified\footnote{The model in~\cite{Ferretti:2014qta} is denoted by M6 in this work and in~\cite{Belyaev:2016ftv}, while the model~\cite{Barnard:2013zea} is denoted by M8.} by the numerical couplings in Table~\ref{allmodels}. Note that, due to the small TCP mass, top loops also give additional sizable contributions to the couplings to gauge bosons (not included in the table, but included in our analysis).

Our goal is to confront the TCP with the existing searches and to propose a new, more sensitive search for such object. We treat the mass $m_a$ and the decay constant $f_a$ of the TCP as free parameters. In composite Higgs UV completions, $f_a$ is related to the composite Higgs decay constant $f_\psi$, entering in the usual alignment parameter $\xi = v^2/f_\psi^2$, by a relative coefficient that was estimated in \cite{Belyaev:2016ftv} and is summarized in Table~\ref{allmodels}. Since bounds on composite Higgs models require $f_\psi\gtrsim800\mbox{ GeV}$, $f_a$ is expected to be naturally of the order of $1\div 2$~TeV.

\begin{table*} \centering
\begin{tabular}{|c|r|r|r|r|r|r|r|r|r|r|r|r|}
 \hline
            & M1  & M2  & M3  & M4  & M5  & M6  & M7  & M8  & M9  & M10  & M11  & M12  \\
             \hline
$K_g$ & -7.2 & -8.7 & -6.3 & -11. & -4.9 & -4.9 & -8.7 & -1.6 & -10. & -9.4 & -3.3 & -4.1\\
$K_W$ & 7.6 & 12. & 8.7 & 12. & 3.6 & 4.4 & 13. & 1.9 & 5.6 & 5.6 & 3.3 & 4.6\\
$K_B$ & 2.8 & 5.9 & -8.2 & -17. & .40 & 1.1 & 7.3 & -2.3 & -22. & -19. & -5.5 & -6.3\\
$C_f$ & 2.2 & 2.6 & 2.2 & 1.5 & 1.5 & 1.5 & 2.6 & 1.9 & .70 & .70 & 1.7 & 1.8\\
$\frac{f_a}{f_\psi}$ & 2.1 & 2.4 & 2.8 & 2.0 & 1.4 & 1.4 & 2.4 & 2.8 & 1.2 & 1.5 & 3.1 & 2.6\\
\hline
\end{tabular}
\caption{Couplings in the twelve models \cite{Belyaev:2016ftv} used as benchmarks.
For the top, several possibilities arise depending on the choice of top partner representation: here, as an illustration, we take the same coupling as for lighter fermions, whose mass arise from bilinear four-fermion interactions.
$f_a/f_\psi$ is an estimate of the ratio between the TCP decay constant $f_a$ and the composite Higgs decay constant $f_\psi$. }\label{allmodels}
\end{table*}

\section{Bounds from existing searches}

\begin{figure}[tbh]
\centering
\includegraphics[width=0.5\textwidth]{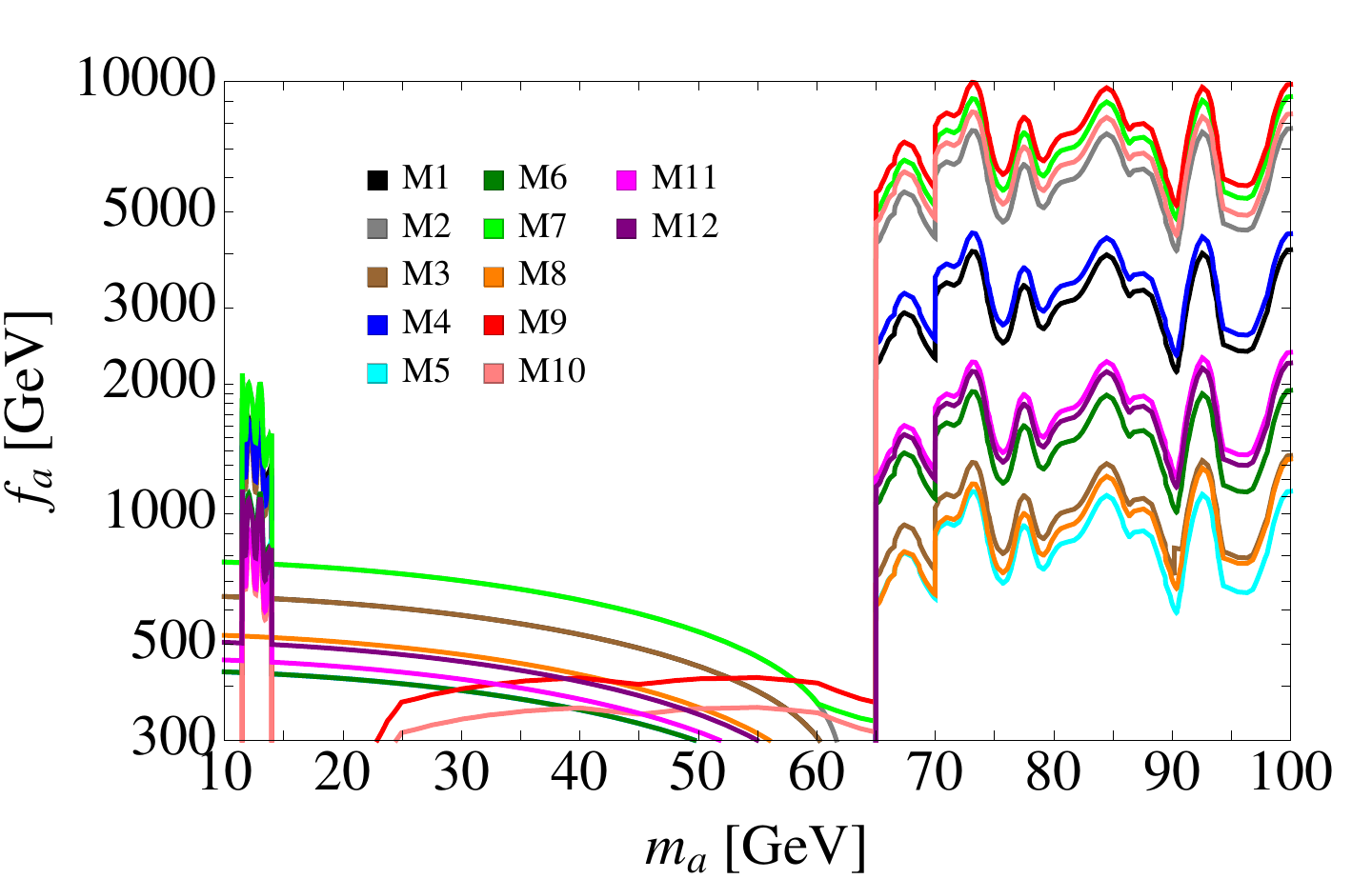}
\caption{Constraints on $f_a$ as a function of $m_a$ for the benchmark models M1 - M12, defined in Table \ref{allmodels}. The bounds arise from di-muon searches \cite{ATLAS:2011cea,Chatrchyan:2012am} in the low mass range, di-photon searches \cite{Aad:2014ioa,CMS-PAS-HIG-17-013} in the higher one, and from the BSM decay width of the Higgs~\cite{Khachatryan:2016vau} below $65$ GeV. We have also indicated the current bounds obtained by adapting the results in \cite{Mariotti:2017vtv} in the region between $20$ and $65$ GeV for the two models (M9 and M10) where they are the strongest. }
\label{bounds}
\end{figure}

Since the TCP is a gauge singlet, its couplings to $Z$ and $W$ are induced by the anomaly and by top loops, thus they are always much smaller than those of a SM Higgs boson. Hence, bounds from all LEP searches for a light Higgs, which are based on $Z$ associated production, are evaded. At hadron colliders the TCP is copiously produced via gluon fusion. However, only very few Tevatron or LHC two-body resonant searches reach down to resonance masses below $\sim 100$ GeV. Relevant bounds arise from Run--I ATLAS  \cite{Aad:2014ioa}  and Run--II CMS \cite{CMS-PAS-HIG-17-013} di-photon searches, which reach down to masses of  $65$~GeV and $70$~GeV, respectively, as well as ATLAS and CMS low-mass di-muon searches \cite{ATLAS:2011cea,Chatrchyan:2012am}, reaching up to $14$~GeV. The bounds on $f_a$ from these searches are shown in Figure~\ref{bounds} \footnote{The 3-$\sigma$ excess at $95.3$ GeV in the Run--II CMS search \cite{CMS-PAS-HIG-17-013} can be seen as a bump, giving a rough estimate of the required value of $f_a$.}, for our models. The bounds are obtained by calculating the leading order TCP production cross section following from the Lagrangian~(\ref{mother}) with PDF set {\tt NNPDF23\_nnlo\_as} {\tt \_0119\_qed}~\cite{Ball:2013hta} and a conservative $k$-factor of 3.3 applied \cite{Ahmed:2016otz}, and using branching ratios into $\gamma\gamma$ and $\mu\mu$ (computed at NLO following Ref.~\cite{Bauer:2017ris}) for the models listed in Table~\ref{allmodels}. Figures on the production cross sections and branching ratios in the twelve sample models are provided in Appendix \ref{app:BRs}.
Resonant di-tau searches reach values of the mass as low as 90 GeV~\cite{CMS-PAS-HIG-16-006,CMS-PAS-HIG-16-037}, however the current bounds are never competitive with the di-photon ones in that range, mainly due to the presence of the Z-peak background.
As noted in Ref.~\cite{Mariotti:2017vtv}, a CMS search looking for boosted $Z'$ in di-jet \cite{CERN-EP-2017-235} may give additional bounds above $50$ GeV.

For other processes, a recent comprehensive review of the existing bounds on ALPs  \cite{Bauer:2017ris} can be directly used to confront TCPs.
Firstly, the one-loop suppression ($1/16\pi^2$) of couplings to vector bosons in the TCP Lagrangian~(\ref{mother}) renders  bounds from vector-boson-fusion or photon-fusion production very weak.
This includes $Z\to a \gamma$ processes and production by photon fusion in Pb-Pb ultra-peripheral collisions \cite{Knapen:2016moh}.
The up-to-date most constraining searches in the mass window between $14$ and $65$~GeV rely on the indirect production via Higgs portal, $h\to a a$. As compared to the generic ALP model discussed in Ref.~\cite{Bauer:2017ris}, the bounds from direct searches are weakened due to the smallness of the $h\rightarrow a a$ branching fraction following from Eq.~(\ref{eq:haa}), and due to the smallness of the $a\rightarrow \gamma\gamma$ and $a\rightarrow \mu\mu$ branching fractions.
Nevertheless, indirect constraints arise from the bounds on the BSM decay width of the Higgs, which is currently below $34\%$ \cite{Khachatryan:2016vau}: as shown in Figure~\ref{bounds}, the lower bound on $f_a$ always falls short of 1 TeV for the models under consideration. \footnote{As the partial width scales with $C_t^4$, choices of larger $C_t$ can lead to bounds on $f_a$ well above 1 TeV. We also remark that this bound will not substantially improve with higher luminosity, with a projected reach of $10\%$ for 3 ab$^{-1}$ \cite{ATL-PHYS-PUB-2014-016}.}
For $m_a<34\ \mbox{ GeV}$ there is also a bound from $h\to Z a$ (following from Eq.(\ref{eq:hZa})), however it turns out even weaker than the Higgs portal one.
Associated $tta$ production may yield a bound on TCPs: using the results of the feasibility study \cite{Casolino:2015cza} at $\sqrt{s}=$ 14 TeV with 3 ab$^{-1}$, which focuses on $a\to bb$ in the mass range between $20$ and $100$ GeV, significant bounds on $f_a$ can be found only for a few models in the low mass end.
Associated $bba$ production yields weaker bounds \cite{Kozaczuk:2015bea}.
Lastly, we should mention that the contribution of the TCP to the anomalous magnetic moment of the muon~\cite{Marciano:2016yhf,Bauer:2017nlg} is also small.
For $m_a=10\mbox{ GeV}$ and $f_a=1\mbox{ TeV}$ it varies from $\Delta a_\mu = -5.7\times 10^{-11}$ for M9 to  $\Delta a_\mu = 2.7\times 10^{-10}$ for M7, the current discrepancy being
$ a_\mu^{\mathrm{exp.}} - a_\mu^{\mathrm{SM}}  = (29.3 \pm 7.6)\times 10^{-10}$.

As shown, the TCP represents an example of a light pseudo-scalar which would evade all existing bounds, while being copiously produced at the LHC in gluon fusion.
Searches in final states from which current bounds arise can be extended in mass range. The low-mass di-muon search \cite{Chatrchyan:2012am} (performed at $\sqrt{s}= 7$~TeV) terminated at $m_a = 14$~GeV, but the first severe physical barrier at higher mass is the di-muon background from Drell-Yan $Z$ production. However, a dedicated low-mass di-muon trigger and a very high invariant mass resolution would be required. \footnote{A LHCb search for a dark photon in di-muon, within the mass range $10$ to $70$~GeV, can be found in Ref.~\cite{Aaij:2017rft}.}

A recent study on inclusive di-photon cross section measurements~\cite{Mariotti:2017vtv} has shown how to extend the low-mass reach of di-photon searches for a generic ALP.  Applying their projected reach to our models we find a nice complementarity between the di-photon channel and our proposal to use the di-tau channel to be discussed below.\footnote{Results of the comparison can be found in Appendix \ref{app:compare}.}

\section{Boosted di-tau searches as a chance to explore the TCP}

As TCP decays to muons and photons have small rates, it is of interest to also look at other final states.
The dominant TCP decay channels are $gg$ and $b\bar{b}$, but both have very large irreducible QCD background.\footnote{Searches for boosted low-mass di-jet resonances are possible, as for instance Ref. \cite{CERN-EP-2017-235}. See also \cite{Chakraborty:2017mbz} for di-tau jets and \cite{Goncalves:2016qhh} for associated top production.} The next-most frequent final state is $\tau^+\tau^-$: sizable rates of a few \% are possible and the models with the lowest rates are the ones with better di-photon reach (see Fig.~\ref{bounds}). Compared to the di-muon channel, the branching ratios are larger by a factor of $\sim m^2_\tau/m^2_\mu\sim 280$.

One of the main challenges for low-mass di-tau resonant searches is the trigger.
The topology that we find most promising is that of a boosted TCP recoiling against an initial state radiation (ISR) jet, and then decaying into $\tau^+ \tau^-$. The boost needs to be sufficient to allow the event to pass the high-level trigger requirement in at least one category (jet, tau or lepton $p_T$) and yet leave enough observable signal. Boosted di-tau pairs have already been considered by CMS~\cite{CMS-PAS-B2G-17-006} in searches for heavy resonances which decay to $hh$, $hZ$, or $hW$. In our case, the mass of the TCP is not known, thus it may not be necessary to require a full reconstruction of the taus with subsequent increase of the systematic uncertainty associated to the procedure. Furthermore, we are interested in light TCPs with a large boost, and thus smaller separation angles between the di-tau decay products can be expected. All decay modes of the di-tau system -- fully hadronic, semi-leptonic, and leptonic -- are potentially interesting. However, for the reasons mentioned above, we focus on the opposite flavor leptonic channel, in which one $\tau$ decays to an electron and the other to a muon. One crucial issue, to be discussed more extensively in Appendix~\ref{app:simulation}, is related to the minimum angular separation $\Delta R_{e\mu}$ between the leptons, since the boosted tau pairs have a small separation angle.
We generate the signal sample $p  p\to a \to \tau^+ \tau^-$ for $m_a=10, 20, \cdots 100$~GeV with up to two jets at the partonic level using MadGraph~\cite{Alwall:2014hca}. We shower and hadronize with Pythia~\cite{Sjostrand:2007bk} and run the fast detector simulation of Delphes~\cite{deFavereau:2013fe} using the standard CMS card after removing the isolation requirement between electrons and muons.
Table~\ref{Tab:signal} shows the value of the signal cross section $\sigma_{\mathrm{prod.}}\times {\mathrm{BR}}_{\tau\tau}$ times the efficiency $\epsilon$ expected for each of the benchmark models with $f_a=1$~TeV after imposing the following requirements:\footnote{For more details on event generation, cut-flows and results, see Appendix \ref{app:simulation}.}
$p_{T\mu}>52$~GeV, $p_{Te}>10$~GeV, $\Delta R_{\mu j}>0.5$, $\Delta R_{e j}>0.5$, $p_{Tj}>150$~GeV,  $\Delta R_{e\mu}<1$,  $m_{e\mu}<100$~GeV. The upper cut on the separation $\Delta R_{e\mu}<1$ is essential in reducing the background from $W$'s~\cite{Khachatryan:2015ywa}, while we do not impose any minimum value yet. (This last issue is discussed below and in Appendix~\ref{app:simulation}.)

\begin{table*}
\begin{center}\begin{tabular}{|c||c|c|c|c|c|c|c|c|c|c|}
  \hline
 $m_a$ & 10 & 20 & 30 & 40 & 50 & 60 & 70 & 80 & 90
   & 100 \\
   \hline\hline
   \text{M1} & 30. & 14. & 9.3 & 6.6 & 5.3 & 3.7 & 3.0 & 2.3 & 1.7 & 1.4 \\
   \text{M2} & 44. & 20. & 13. & 9.5 & 7.7 & 5.4 & 4.4 & 3.2 & 2.4 & 2.0 \\
   \text{M3} & 26. & 12. & 8.4 & 6.1 & 5.0 & 3.6 & 2.9 & 2.2 & 1.6 & 1.4 \\
   \text{M4} & 28. & 11. & 6.1 & 3.8 & 2.9 & 1.9 & 1.5 & 1.1 & 0.80 & 0.67 \\
   \text{M5} & 14. & 6.3 & 4.2 & 3.0 & 2.4 & 1.7 & 1.4 & 1.0 & 0.74 & 0.63 \\
   \text{M6} & 14. & 6.3 & 4.2 & 3.0 & 2.4 & 1.7 & 1.4 & 1.0 & 0.74 & 0.63 \\
   \text{M7} & 44. & 20. & 13. & 9.5 & 7.7 & 5.4 & 4.4 & 3.2 & 2.4 & 2.0 \\
   \text{M8} & 4.0 & 2.1 & 1.8 & 1.6 & 1.6 & 1.3 & 1.2 & 0.96 & 0.76 & 0.69 \\
   \text{M9} & 8.3 & 3.1 & 1.6 & 0.95 & 0.70 & 0.47 & 0.36 & 0.26 & 0.19 & 0.16 \\
   \text{M10} & 8.1 & 3.0 & 1.6 & 0.95 & 0.70 & 0.46 & 0.36 & 0.26 & 0.19 & 0.16 \\
   \text{M11} & 9.4 & 4.7 & 3.5 & 2.8 & 2.4 & 1.8 & 1.5 & 1.2 & 0.87 & 0.74 \\
   \text{M12} & 13. & 6.4 & 4.7 & 3.6 & 3.1 & 2.3 & 1.9 & 1.4 & 1.1 & 0.92 \\
  \hline
\end{tabular}\end{center}
\caption{The values of $\sigma_{\mathrm{prod.}} \times BR_{\tau\tau}\times \epsilon$ in fb for $f_a = 1$~TeV for each of the models defined in Table~\ref{allmodels}.  The  main backgrounds are di-top and single-top
(59.2~fb), $Z/\gamma^*$ (24.7~fb), and di-bosons (11.0~fb).}
\label{Tab:signal}
\end{table*}

\begin{figure}[tbh]
\centering
\includegraphics[width=0.5\textwidth]{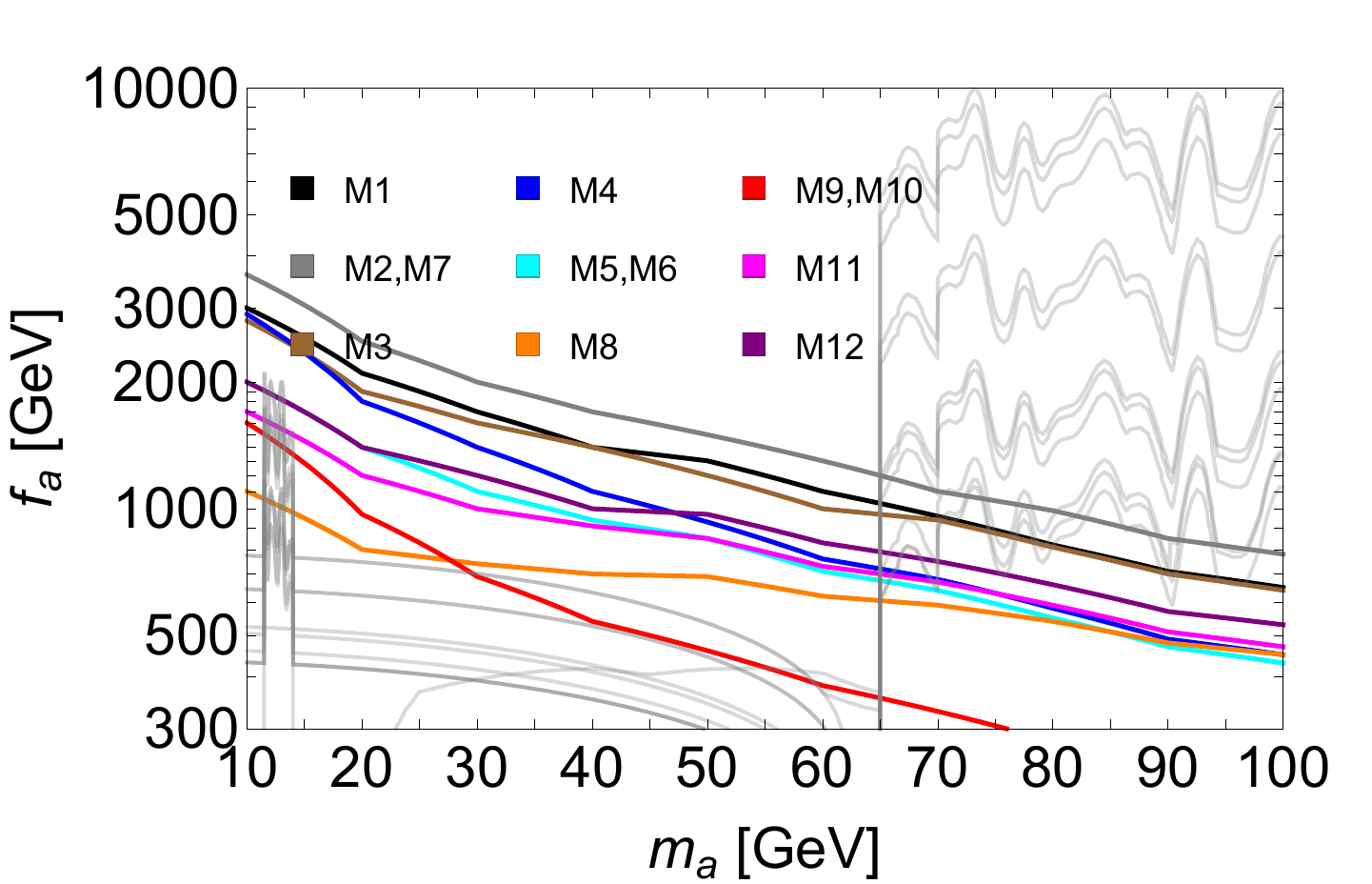}
\caption{Values of $f_a$ for models M1 - M12 for which $\mathcal{Z} \equiv S/\sqrt{B+\delta^2 B^2} = 3$ in the proposed di-tau search after an integrated Luminosity of 300 fb$^{-1}$. We assume a systematic error on the background $\delta = 1 \%$. Shown in grey are the current bounds as of Figure~\ref{bounds}.}
\label{ditau}
\end{figure}

The leading irreducible SM backgrounds are  $t \bar t +$ single top, $\gamma^*/  Z$ and $VV$, the last one being mainly $W^+ W^-$. Our simulation of these backgrounds yields $\sigma\times \epsilon = 59$, $25$ and $11$~fb, respectively, after imposing the same cuts as above. As discussed in Appendix~\ref{app:simulation}, we expect the reducible backgrounds of single vector boson+fakes and QCD to be sub-leading (of the order of a few fb) in the $(\mu, e)$ channel. Note that we do not require full reconstruction of the tau momentum.

\begin{figure}[tbh]
\centering
\vspace{-13pt}
\includegraphics[width=0.5\textwidth]{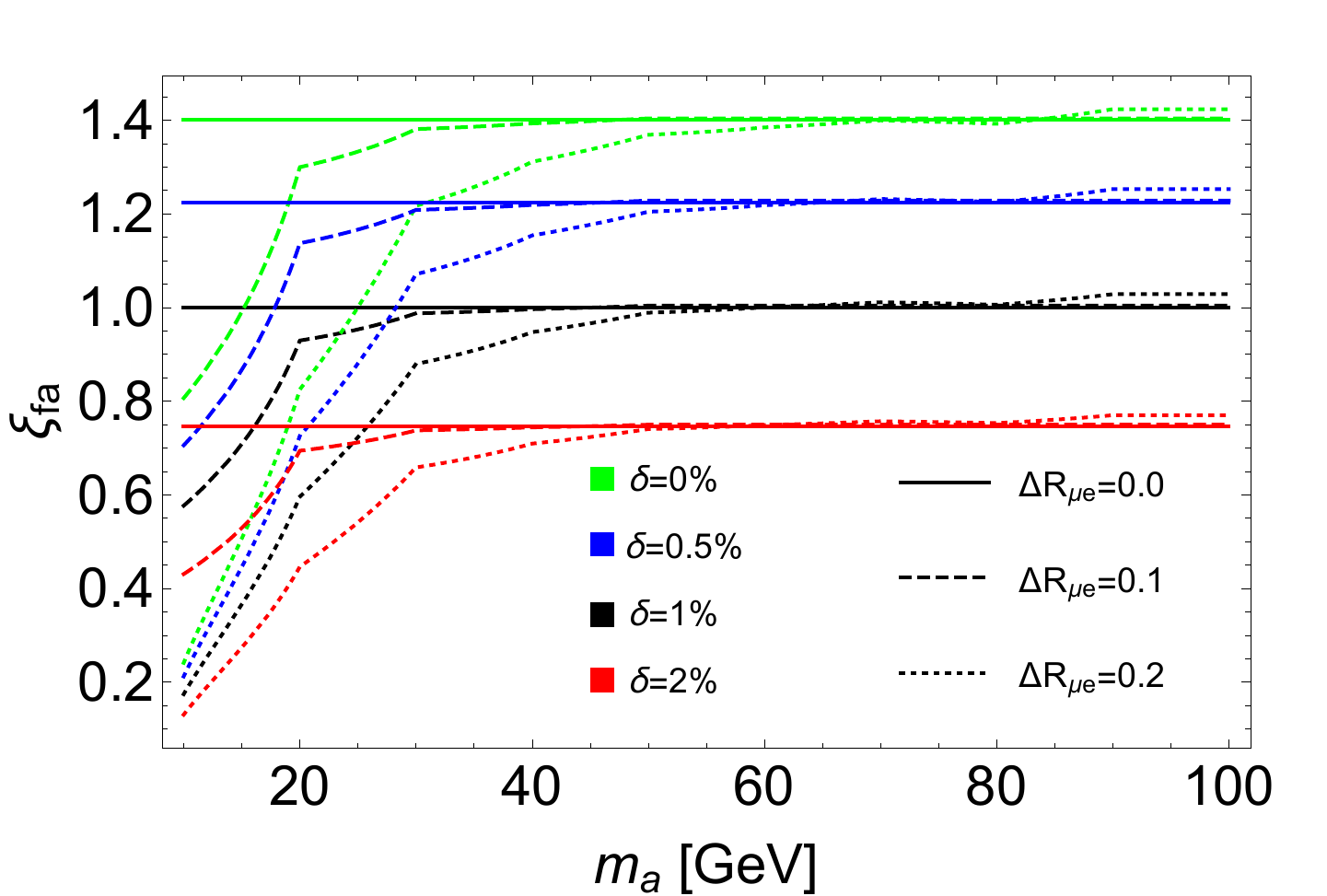}
\caption{Relative change $\xi_{f_a}\equiv f_a^{\delta,\Delta R_{e\mu}}  / f_a^{1\%, 0}$ in the projected bounds on $f_a$ with 300 fb$^{-1}$ of data.
We plot the relative change against the baseline presented in Fig.~\ref{ditau} for different values of systematic uncertainties $\delta = 0, 0.5, 1,$ and $2 \%$ (green, blue, black, and red) and choosing three different separation cuts $\Delta R_{e\mu} > 0$ (solid) $0.1$ (dashed) and $0.2$ (dotted) respectively.}
\label{ratios}
\end{figure}

To be able to estimate the reach of Run II+III at the LHC, it is essential to know the systematic uncertainties because we are in a situation where the signal over background ratio is small, $S/B\ll 1$, as can be seen from Table~\ref{Tab:signal}. Since a search of this type has not been done by the experimental collaborations, we cannot reliably quantify the systematic uncertainties yet. Data driven techniques can certainly be used to reduce the systematic uncertainties on the different backgrounds. For the lepton identification, the systematic uncertainties typically amount to 1\%~\cite{Khachatryan:2015ywa}, while we do not require tau identification, which would increase the systematics to 10--20\%. To assess the feasibility of the analysis, we will include the systematic uncertainty $\delta$ in the significance $\mathcal{Z}$ according the approximate formula
\begin{equation}
\mathcal{Z}= \frac{S}{\sqrt{B + \delta^2 B^2}}\,,
\label{sigmafom}
\end{equation}
where $S$ and $B$ are the number of signal and background events at a given integrated luminosity and $\delta$ the relative systematic error on the background.
Eq. (\ref{sigmafom}) works quite well in the regime of interest for this work when compared to the more general treatment in~\cite{Cowan:2010js}.
A projection of the bound on $f_a$ for the various models after 300 fb$^{-1}$ integrated luminosity is shown in Figure~\ref{ditau}, including an estimated systematic uncertainty of 1\%.
We can see that for all models with exception of M9 and M10, the boosted di-tau search we propose can probe the mass range of $10-70$ GeV with integrated luminosity below $300$ fb$^{-1}$.

n Figure~\ref{ratios} we show the relative change in the projected bound on $f_a$ if a minimum $\Delta R_{e\mu}$ cut of $0.1$ or $0.2$ is imposed as well as its dependence on a change in systematic uncertainty from 1\% to 0, 0.5, and 2\%. These changes apply to all models in a universal way. We can see a loss of sensitivity for masses below $\sim 30 - 40$ GeV when raising $\Delta R_{e\mu}$, while above this mass range the search is barely affected. Thus, being able to remove or reduce the minimum separation angle is important for the lowest mass region, as long as it does not imply an increase in the systematic errors.

The plot also clearly shows the importance of controlling the systematic uncertainties to a level close to 1\%. The latter values are what CMS and ATLAS typically require for opposite sign leptons~\cite{Khachatryan:2015ywa} in current searches.

\begin{figure}[tbh]
	\centering
	\hspace{-20pt}
	\includegraphics[width=0.5\textwidth]{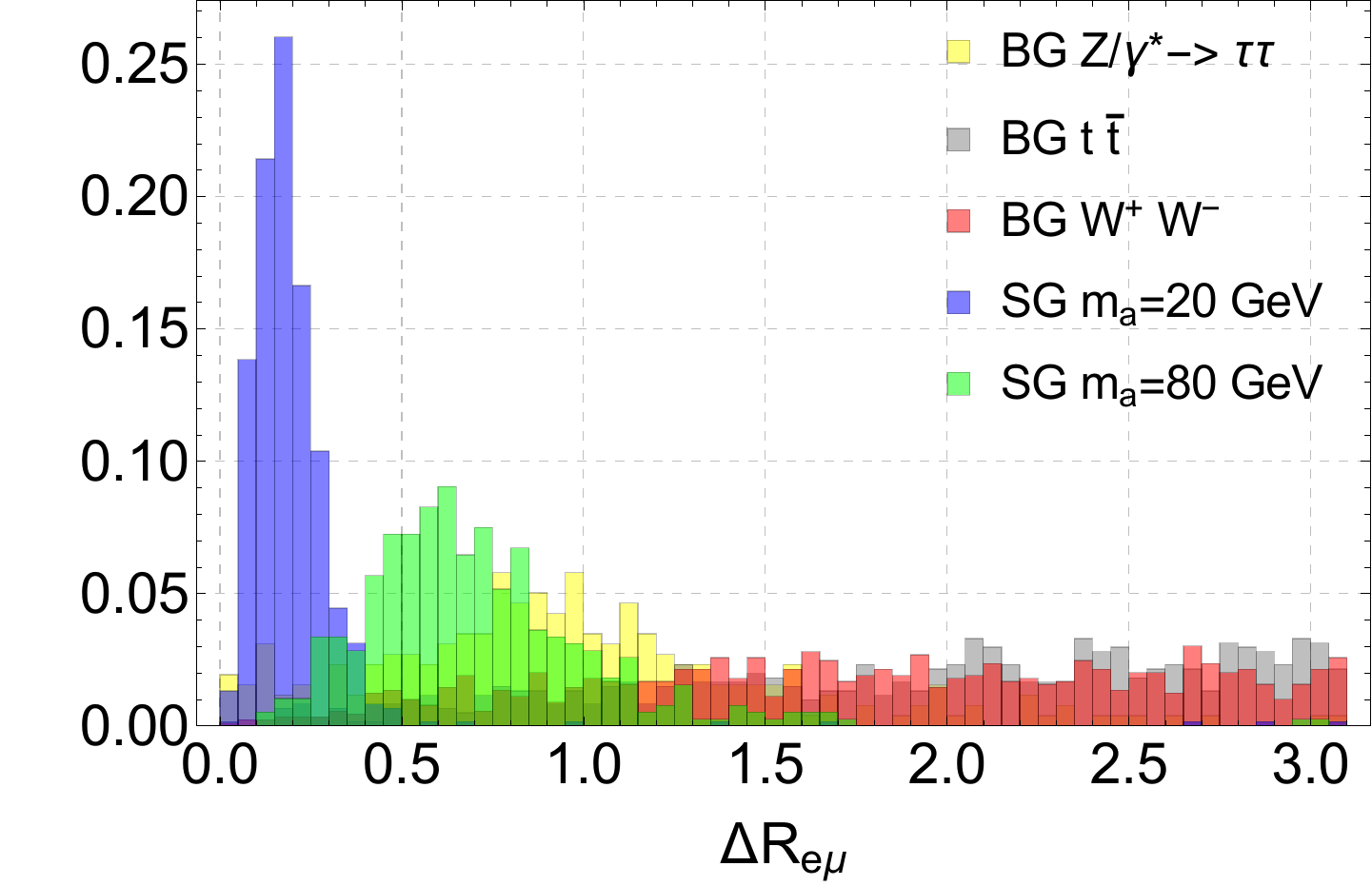}
	\caption{Angular separation ($\Delta R_{e\mu}$) between the two leptons for two signal (SG) masses (20 GeV  and 80 GeV) compared to the most relevant backgrounds (BG). Small separation angles can be a good discriminant particularly for low masses. }
	\label{dR}
\end{figure}

It is possible to improve sensitivity by imposing variable cuts on the invariant mass $m_{e\mu}$ and particularly on the angular separation $\Delta R_{e\mu}$ of the lepton pair depending on the mass range of interest. For guidance we show in Figure~\ref{dR} the kinematic distribution of $\Delta R_{e\mu}$ for the most relevant backgrounds and the signal with  $m_a = 20$ and $80$~GeV before imposing the $m_{e\mu}<100$~GeV and  $\Delta R_{e\mu}<1$ cuts.

As mentioned above, fully or semi-hadronic decays of the di-tau system may also be testable by designing appropriate di-tau jet algorithms.
For the semi-hadronic case, a sophisticated isolation procedure has been used by CMS for boosted Higgs tagging in the di-tau channel. However, large systematic uncertainties, or the order of 20-30\%~\cite{CMS-PAS-B2G-17-006}, are introduced due to the modified isolation and tau-identification procedures. Furthermore, the signal we are interested in features smaller separation between the two taus, thus a better performance may be achieved by a dedicated identification procedure. For instance, the technique of ``mini-isolation'' proposed in Ref.~\cite{Rehermann:2010vq} may be adapted to this case, although this is beyond the scope of this paper.
For the fully hadronic case, preliminary studies in Refs \cite{Katz:2010iq,Conte:2016zjp} show that a good discrimination between di-tau jets and single tau or QCD jets can be achieved using sub-jet variables. However, a correct estimate of the background (especially from QCD) can only be done with data driven techniques, thus we do not attempt to quantify the sensitivity of these channels.

\section{Conclusions}
The search for new resonances at the LHC continues, and many searches for high-mass resonances are being performed. Nevertheless, complementary searches for lower-mass resonances which have evaded current constraints must not be forgotten.
We observe that light pseudo-scalars in the mass regime between $14$ and $65$ GeV can be copiously produced at the LHC while avoiding current experimental constraints.

We propose to search for boosted di-tau resonances, produced via gluon fusion, that can effectively cover this open window.
We test this strategy on a set of twelve benchmark models of composite Higgs with top partial compositeness, which have a simple gauge-fermion underlying description.
Low mass di-photon searches effectively cover masses above $65$ GeV. Extending the di-photon search to lower masses is challenging due to triggers (but potential solutions have been presented ~\cite{Mariotti:2017vtv}), while resuming low-mass di-muon resonant searches and extending them to higher masses is  possible but challenging due to increased muon $p_T$ trigger thresholds.
The boosted di-tau search we propose allows to access the open window below $65$ GeV, and, for some models, it can be competitive with the di-photon channel at higher masses.

\section*{Acknowledgements}
We thank Hwidong Yoo for helpful comments and suggestions on lepton-isolation and di-muon searches, Suzanne Gascon-Shotkin for comments on di-photon and di-tau searches. We thank Alberto Mariotti, Diego Redigolo and Filippo Sala for discussions about the relation of their proposed search to ours. We thank Matthias Neubert for email communications.  GF and GC thank for hospitality the IBS CTPU, where part of this work was performed. TF was supported by IBS under the project code IBS-R018-D1. GF is supported in part by a grant from the Wallenberg foundation.
GC acknowledges partial support from the Institut Franco-Suedois (project T{\"o}r) and the Labex Lyon Institute of the Origins - LIO. HS has received funding  from  the  European  Research Council (ERC) under the European Union's Horizon 2020 research and innovation programme (grant agreement No 668679).

\appendix
\section*{Appendices}

\section{Details of the models}\label{app:models}

A light Higgs may emerge as the pseudo-Nambu-Goldstone boson (pNGB) of a spontaneously broken global symmetry of a new strongly interacting sector. Most of the developments in these models have been achieved by use of effective descriptions. However, as shown in~\cite{Belyaev:2016ftv}, the knowledge of the underlying theory (a partial UV completion) based on gauge-fermion models allows for specific, and testable, predictions.

A particular minimalist class of UV completions was found in~\cite{Ferretti:2013kya}. In such models, dangerous leptoquark pNGBs are avoided by having two distinct cosets, one associated with color and the other with electroweak quantum numbers. The presence of two cosets also requires the presence of two species of underlying fermions in the theory, $\psi$ and $\chi$. Imposing further dynamical assumptions, one is left with twelve models, whose properties are summarized in Tab.~\ref{tab: Models}. The details of the models have been extensively explored in Ref.~\cite{Belyaev:2016ftv}, so here we will simply recall their main features.

\begin{table*}[h]
\begin{center}
\begin{tabular}{|c|lllcc|c|}
\hline
Coset&HC&$\psi$&$\chi$&$-q_\chi/q_\psi$&$Y_\chi$&Model\\
\hline
\hline
\multirow{4}{*}{$\dfrac{SU(5)}{SO(5)}\times \dfrac{SU(6)}{SO(6)}$}&$SO(7)$&\multirow{2}{*}{$5\times \mathbf{F}$}&\multirow{2}{*}{$6\times \textbf{Spin}$}&$5/6$&\multirow{2}{*}{1/3}&M1\\
&$SO(9)$&&&$5/12$&&M2\\
\cline{2-6}
&$SO(7)$&\multirow{2}{*}{$5\times \textbf{Spin}$}&\multirow{2}{*}{$6\times \text{F}$}&$5/6$&\multirow{2}{*}{2/3}&M3\\
&$SO(9)$&&&$5/3$&&M4\\
\hline
\hline
&&&&&&\\[-2ex]
$\dfrac{SU(5)}{SO(5)}\times \dfrac{SU(6)}{Sp(6)}$&$Sp(4)$&$5\times \textbf{A}_2$&$6\times \textbf{F}$&$5/3$&1/3&M5\\[2ex]
\hline
\hline
&&&&&&\\[-2ex]
\multirow{2}{*}{$\dfrac{SU(5)}{SO(5)}\times \dfrac{SU(3)^2}{SU(3)}$}&$SU(4)$&$5\times \textbf{A}_2$&$3\times (\textbf{F},\overline{\textbf{F}})$&$5/3$&\multirow{2}{*}{1/3}&M6\\
&$SO(10)$&$5\times \textbf{F}$&$3\times (\textbf{Spin},\overline{\textbf{Spin}})$&$5/12$&&M7\\[1ex]
\hline
\hline
&&&&&&\\[-2ex]
\multirow{2}{*}{$\dfrac{SU(4)}{Sp(4)}\times \dfrac{SU(6)}{SO(6)}$}&$Sp(4)$&$4\times \textbf{F}$&$6\times \textbf{A}_2$&$1/3$&\multirow{2}{*}{2/3}&M8\\
&$SO(11)$&$4\times \textbf{Spin}$&$6\times \textbf{F}$&$8/3$&&M9\\[1ex]
\hline
\hline
&&&&&&\\[-2ex]
\multirow{2}{*}{$\dfrac{SU(4)^2}{SU(4)}\times \dfrac{SU(6)}{SO(6)}$}&$SO(10)$&$4\times (\textbf{Spin},\overline{\textbf{Spin}})$&$6\times \textbf{F}$&$8/3$&\multirow{2}{*}{2/3}&M10\\
&$SU(4)$&$4\times (\textbf{F},\overline{\textbf{F}})$&$6\times \textbf{A}_2$&$2/3$&&M11\\[1ex]
\hline
\hline
&&&&&&\\[-2ex]
$\dfrac{SU(4)^2}{SU(4)}\times \dfrac{SU(3)^2}{SU(3)}$&$SU(5)$&$4\times (\textbf{F},\overline{\textbf{F}})$&$3\times (\textbf{A}_2,\overline{\textbf{A}_2})$&$4/9$&2/3&M12\\[2ex]
\hline
\end{tabular}
\caption{\label{tab: Models} The first column shows the EW and color cosets, respectively. The $-q_\chi/q_\psi$ column indicates the ratio of charges of the fermions under the non anomalous $U(1)$ combination. HC is the confining hyper-color gauge group. $\textbf{F}$ and $\textbf{A}_2$ denote the fundamental and anti-symmetric representation of HC.}
\end{center}
\end{table*}

In order to study the low energy degrees of freedom of the present models we use the formulation of chiral perturbation theory. We parameterize relevant degrees of freedom as
\begin{equation}
\Sigma_r=\text{exp}\left[i2\sqrt{2}c_5 \frac{\pi^a_r T^a_r}{f_r}\right].\Sigma_{0,r}\quad \text{and}\quad \Phi_r=\text{exp}\left[ic_5 \frac{a_r}{f_{a_r}}\right]\,,
\end{equation}
where $r=\psi$ or $\chi$.
The meson matrices $\Sigma_r$ are associated with electroweak and color cosets. The number of fields $\pi_r^a$, associated with the broken generators of the non-Abelian symmetries, is model dependent. The  matrix $\Sigma_{0,r}$ is the gauge-preserving vacuum. The meson matrix $\Sigma_\psi$ contains a Higgs in the custodial representation on top of extra scalar multiplets. The coefficient $c_5$ is $\sqrt{2}$ when the $\psi$ irrep is real and $1$ otherwise.

While the meson matrices $\Sigma_r$ have a different structure for each coset, the Abelian terms $\Phi_r$ characterize a universal feature of all the models, i.e. the presence of two pseudo-scalars in the spectrum. They are associated with the freedom to preform a chiral rotation in the underlying fermions $\psi$ and $\chi$, or, in other words, with the presence of the Abelian symmetries $U(1)_{\psi,\chi}$. These symmetries are spontaneously broken by the respective condensates, and explicitly broken by the underlying fermion masses and the gauging of the SM symmetries (via anomalies). On top of this, one combination has an anomaly with the new strong dynamics gauge bosons. The charges for the non-anomalous $U(1)$ are given in Tab.~\ref{tab: Models}. For each model we can, therefore, find the linear combinations of $a_{\psi,\chi}$ that are anomaly-free and anomalous, i.e.
\begin{equation}
\tilde{a}=\frac{q_\psi f_{a_\psi}a_\psi+q_\chi f_{a_\chi}a_\chi}{\sqrt{q_\psi^2 f_{a_\psi}^2+q_\chi^2 f_{a_\chi}^2}}\,\quad\mbox{and}\quad
\tilde{\eta}^\prime=\frac{q_\psi f_{a_\psi}a_\chi-q_\chi f_{a_\chi}a_\psi}{\sqrt{q_\psi^2 f_{a_\psi}^2+q_\chi^2 f_{a_\chi}^2}}
\end{equation}
respectively. These combinations are, in general, not the physical mass eigenstates. Ref.~\cite{Belyaev:2016ftv} studied the phenomenology of these two states in the case where they are both heavy (above 500 GeV). In this article, we are only interested in the limit where one of them is light: this can only happen for the pNGB associated to the non-anomalous U(1), i.e. $\tilde{a}$, while the other one is massive and decouples. Thus, the light mass eigenstate $a$ coincides with the non-anomalous $\tilde{a}$.

The couplings of the light pNGB to gauge bosons and fermions can be expressed as (we match the notation used in~ (\ref{mother}) with the one used in Ref.~\cite{Belyaev:2016ftv}):
\begin{equation}
K_A=c_5\frac{C_A^\psi q_\psi+C_A^\chi q_\chi}{\sqrt{q_\psi^2 + q_\chi^2}}\quad \text{and}\quad C_t=c_5\frac{n_\psi q_\psi+n_\chi q_\chi}{\sqrt{q_\psi^2 + q_\chi^2}}\,,
\end{equation}
with $A=g,\, W,\, B$, and we have normalized the coefficients in such a way to render them independent on the normalization of the charges. The numerical values for each model are given in Tab.~\ref{allmodels}. The $C_A^{\psi,\chi}$ are WZW coefficients of the anomaly terms between $U(1)_{\psi,\chi}$ and the SM gauge group. These are completely determined by the underlying fermionic representation. We always have $C_W^\psi=C_B^\psi =d_\psi$ (dimension of the $\psi$ irrep under the hyper-color group) for complex and real irreps and $ C_W^\psi=C_B^\psi =d_\psi/2$ for pseudo-real ones. For the fermion $\chi$ we have $C_G^\chi=d_\chi$ and $C_{B}^\chi=6Y_\chi^2 d_\chi$ for all irreps. The decay constant in~\eqref{mother} is related to the decay constants of the two condensates as
\begin{equation}
f_a=\sqrt{\frac{q_\psi^2 f_{a_\psi}^2+q_\chi^2 f_{a_\chi}^2}{q_\psi^2 + q_\chi^2}}\,.
\end{equation}
To quantify the amount of fine-tuning, it is necessary to give an estimate of this decay constant with respect to the one entering the Higgs sector. To do so, we relate the decay constants in the abelian and non-abelian sectors by use of large-$N$ estimates: $f_{a_\psi}=\sqrt{N_\psi}f_\psi$ and $f_{a_\chi}=\sqrt{N_\chi}f_\chi$. This leads to:
\begin{equation}
\frac{f_a}{f_\psi}= \sqrt{\left(N_\psi+N_\chi \frac{q_\chi^2}{q_\psi^2} \frac{f_\chi^2}{f_\psi^2}\right)/\left(1+\frac{q_\chi^2}{q_\psi^2} \right)}\,.
\end{equation}
The ratio $f_\psi/f_\chi$ can be estimated based on the MAC hypothesis~\cite{Belyaev:2016ftv}, leading to
\begin{eqnarray}
f_\psi/f_\chi&=&\{1.4,\,0.75,\,0.73,\, 1.3,\, 2.8,\, 1.9,\, \phantom{\}} \nonumber \\
&&\phantom{\{}0.58, \, 0.38,\, 2.3,\, 1.7,\, 0.52, \, 0.38\}
\end{eqnarray}
for models M1, \ldots, M12.

\begin{table}
\begin{center}\begin{tabular}{|l|c|c|c|c|c|}
  \hline
$(n_\psi, n_\chi)$ & $(\pm 2, 0)$ & $(0, \pm 2)$  & $(4,2)$  & $(-4,2)$   \\
                           &                      &                      & or $(2,4)$  & or $(2,-4)$  \\
  \hline\hline
 M1 & $\pm 2.2$ & $\mp 1.8$ & $-1.4$ & $5.8$  \\
 M2 & $\pm 2.6$ & $\mp 1.1$ & $0.44$ & $4.8$\\
 M3 & $\pm 2.2$ & $\mp 1.8$ & $2.5$ & $-6.2$ \\
 M4 & $\pm 1.5$ & $\mp 2.4$ & $0.49$ & $-5.3$ \\
 M5 & $\pm 1.5$ & $\mp 2.4$ & $-3.4$ & $6.3$ \\
 M6 & $\pm 1.5$ & $\mp 2.4$ & $-3.4$ & $6.3$ \\
 M7 & $\pm 2.6$ & $\mp 1.1$ & $0.44$ & $4.8$ \\
 M8 & $\pm 1.9$ & $\mp 0.63$ & $3.2$ & $-4.4$ \\
 M9 & $\pm 0.70$ & $\mp 1.9$ & $-0.47$ & $-3.3$ \\
 M10 & $\pm 0.70$ & $\mp 1.9$ & $-0.47$ & $-3.3$ \\
 M11 & $\pm 1.7$ & $\mp 1.1$ & $2.2$ & $-4.4$ \\
 M12 & $\pm 1.8$ & $\mp 0.81$ & $2.8$ & $-4.5$ \\
  \hline
\end{tabular}\end{center}
\caption{Values of $C_t$ for the various possible top partner assignment. For the last two columns, the values correspond to $(\pm 4, 2)$ for $\psi \psi \chi$ models ($Y_\chi = 2/3$), and $(2, \pm 4)$ for $\psi \chi \chi$ models ($Y_\chi = 1/3$). In the paper we used the assignment $(2,0)$ which is also common to all other fermions.}
\label{tab:Ct}
\end{table}

We finally want to comment on the couplings to fermions. For the light quarks and leptons, we assume that they couple to the strong dynamics via bi-linear four fermion interactions only involving $\psi$'s. This is mainly due to the fact that it is impossible to generate enough partners to make all SM fermions partially composite. Thus, the couplings are obtained by setting $(n_\psi, n_\chi) = (2,0)$ in the above expressions.
For the top, the coupling crucially depends on the charges of the top partners: if the bound state is made of $\psi \psi \chi$, the possible assignments are  $(n_\psi, n_\chi) = (\pm 2, 0)$, $(0, \pm 2)$, $(\pm 4, 2)$, while for $\psi \chi \chi$ one has $(n_\psi, n_\chi) = (\pm 2, 0)$, $(0, \pm 2)$, $(2, \pm 4)$. The values of the couplings $C_t$ for all models and all assignments are reported in Table~\ref{tab:Ct}.
In the main text, we present results for the case $(2,0)$, which gives the same coupling to all SM fermions. This is only a representative case. Note that the bounds from other searches also depend on this choice.

\subsection{Loop calculation for the $h\to aa$ and $h \to Z a$ decays}\label{app:loops}

\begin{figure}[tbh]
\centering
\includegraphics[width=0.5\textwidth]{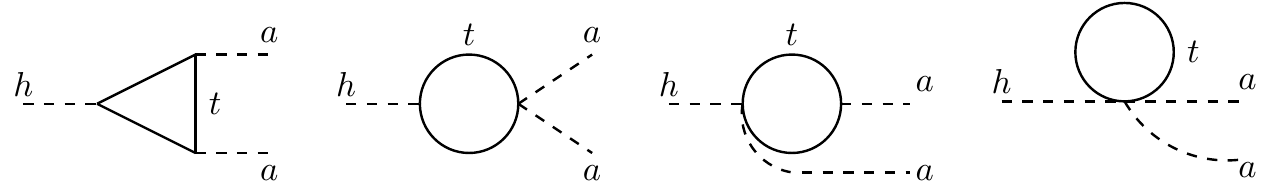}
\caption{Top loop diagrams contributing to the effective coupling $h \to aa$.}
\label{fig:diag}
\end{figure}

The coupling of the Higgs to two TCPs is generated mainly by loops of top quarks (the contribution of lighter fermions being suppressed by the mass, while the gauge contribution is suppressed by the small anomalous couplings). The relevant vertices can be read off by expanding the spurion term $   - m_t(h)\ \mathrm{e}^{i C_t a/f_a} \bar\Psi_{tL} \Psi_{tR} + \mathrm{h.c.}$ as follows:
\begin{eqnarray}
&&- m_t\ \bar\Psi_t \Psi_t - i \frac{C_t m_t}{f_a} a\ \bar\Psi_t \gamma^5 \Psi_t + \frac{C_t^2 m_t}{2 f_a^2} a^2\ \bar\Psi_t \Psi_t\\
&&- \frac{m_t}{v} \kappa_t h \left( \bar\Psi_t \Psi_t + i \frac{C_t}{f_a} a\ \bar{\Psi}_t \gamma^5 \Psi_t - \frac{C_t^2}{2 f_a^2} a^2\ \bar\Psi_t \Psi_t \right) + \dots  \nonumber
\end{eqnarray}
where the Higgs coupling is defined as
\begin{equation}
\frac{m_t}{v} \kappa_t = \left. \frac{\partial m_t (h)}{\partial h} \right|_{h \to 0}\,,
\end{equation}
so that $\kappa_t$ encodes the deviations from the SM coupling $m_t/v$. The Lagrangian above allows for four diagrams, depicted in Figure~\ref{fig:diag}.
The last three contain a quadratic divergence, that vanishes once they are summed. Thus, we are left with a log divergence that contributes to the amplitude as:
\begin{equation}
i \Sigma = - i \frac{C_t^2 m_t^2}{v f_a^2} \kappa_t \frac{3}{8 \pi^2} \log \frac{\Lambda^2}{m_t^2}\ (p_h^2 - p_{a1}^2 - p_{a2}^2) + \mbox{finite}
\end{equation}
where $p_h$ and $p_{ai}$ are the four-momenta of the Higgs and of the two TCPs respectively. The operator in Eq. (\ref{eq:haa}) matches the divergent part of the above amplitude. Note also that the result differs from the one in Ref.~\cite{Bauer:2017ris} by a factor of $1/4$ while having the same form: the two calculations indeed refer to two different models, as in Ref.~\cite{Bauer:2017ris} only a derivative coupling  is considered.

The calculation of the $h \to Z a$ vertex is the same as in Ref.~\cite{Bauer:2016zfj}, except for the modifications of the Higgs couplings $\kappa_t$ and $\kappa_V$ that prevent the cancellation of the log divergences.

\subsection{Cross sections, widths and branching ratios}\label{app:BRs}

\begin{figure}[h!tb]
\centering
\includegraphics[width=0.5\textwidth]{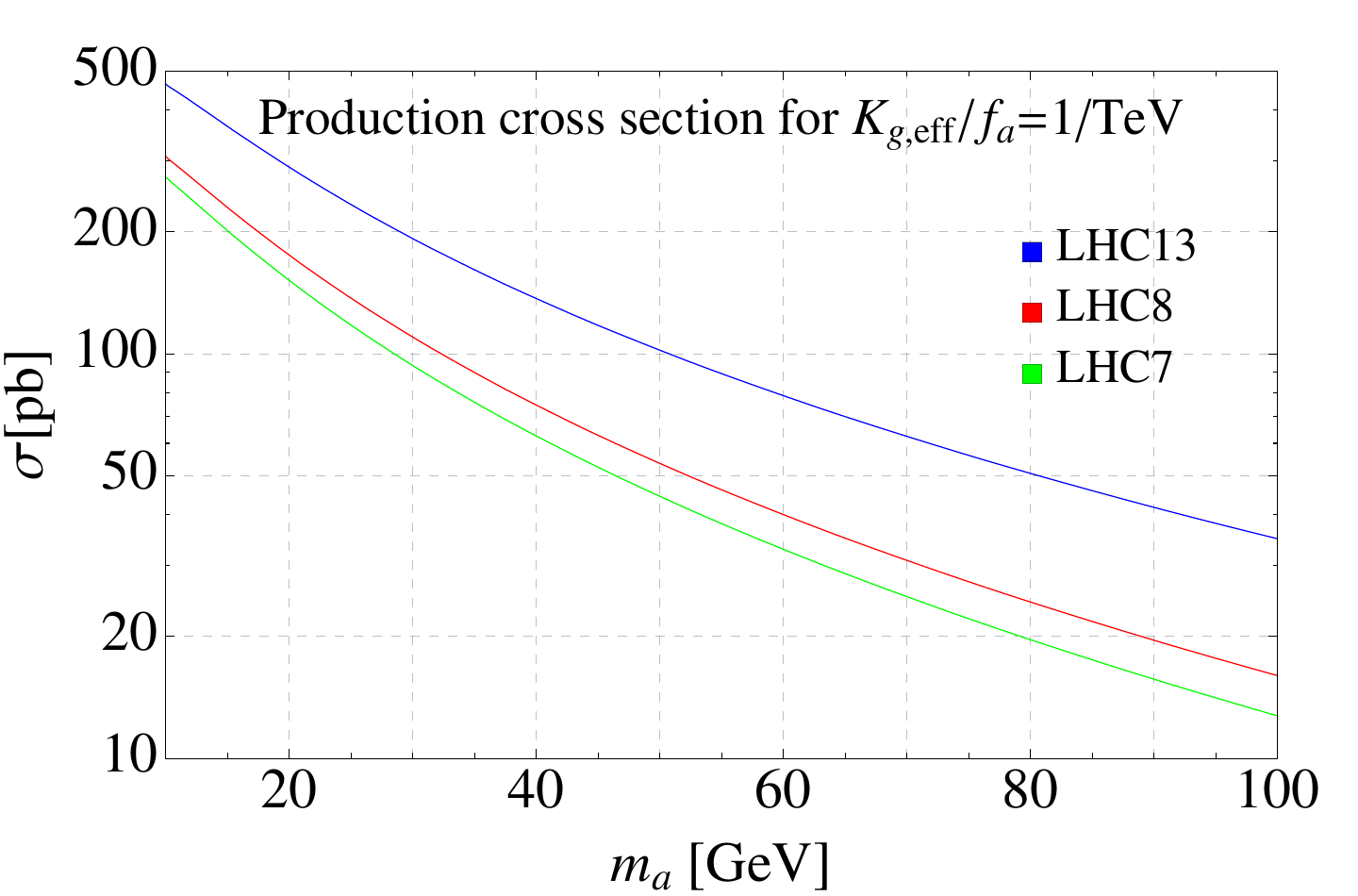}
\caption{Production cross section of $a$ for LHC with $\sqrt{s}=7,8,13$~TeV, for $K_{g,\rm eff}/f_a=1/\mbox{TeV}$.}
\label{fig:prod}
\end{figure}

\begin{figure*}[h!tb]
\centering
\begin{tabular}{cc}
\hspace{-10pt}\includegraphics[width=0.5\textwidth]{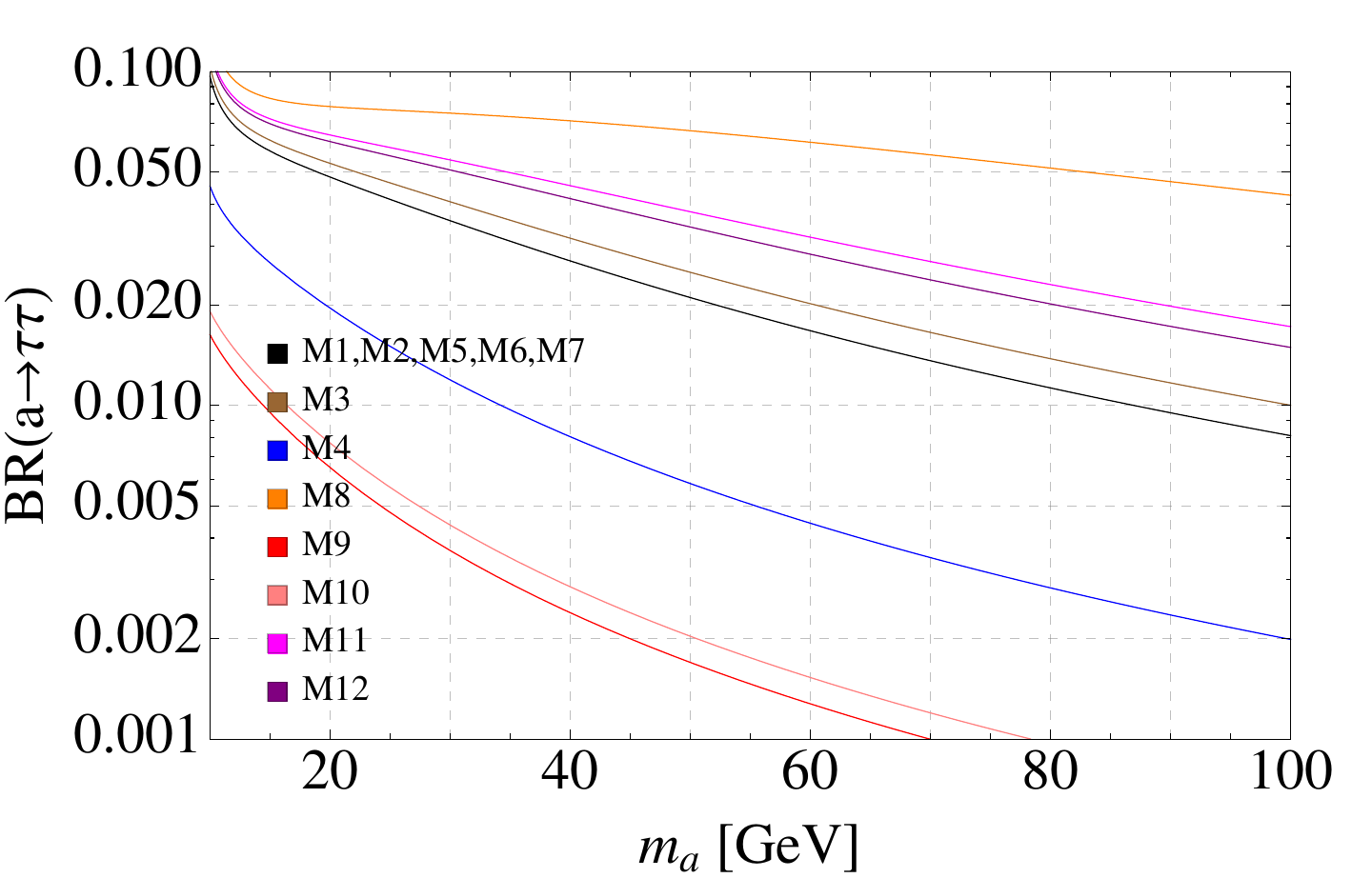}&
\includegraphics[width=0.5\textwidth]{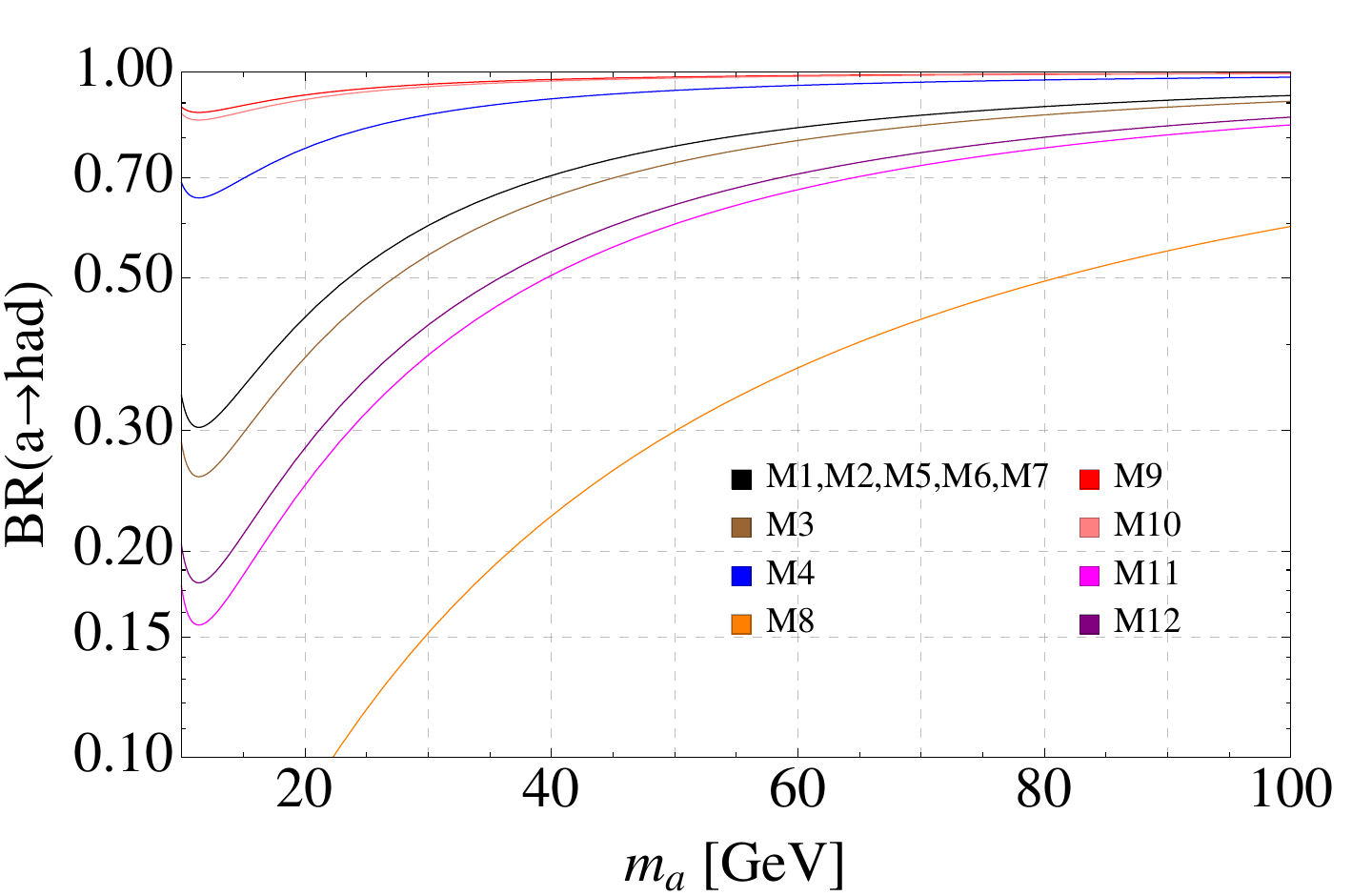}\\
\hspace{-10pt}\includegraphics[width=0.5\textwidth]{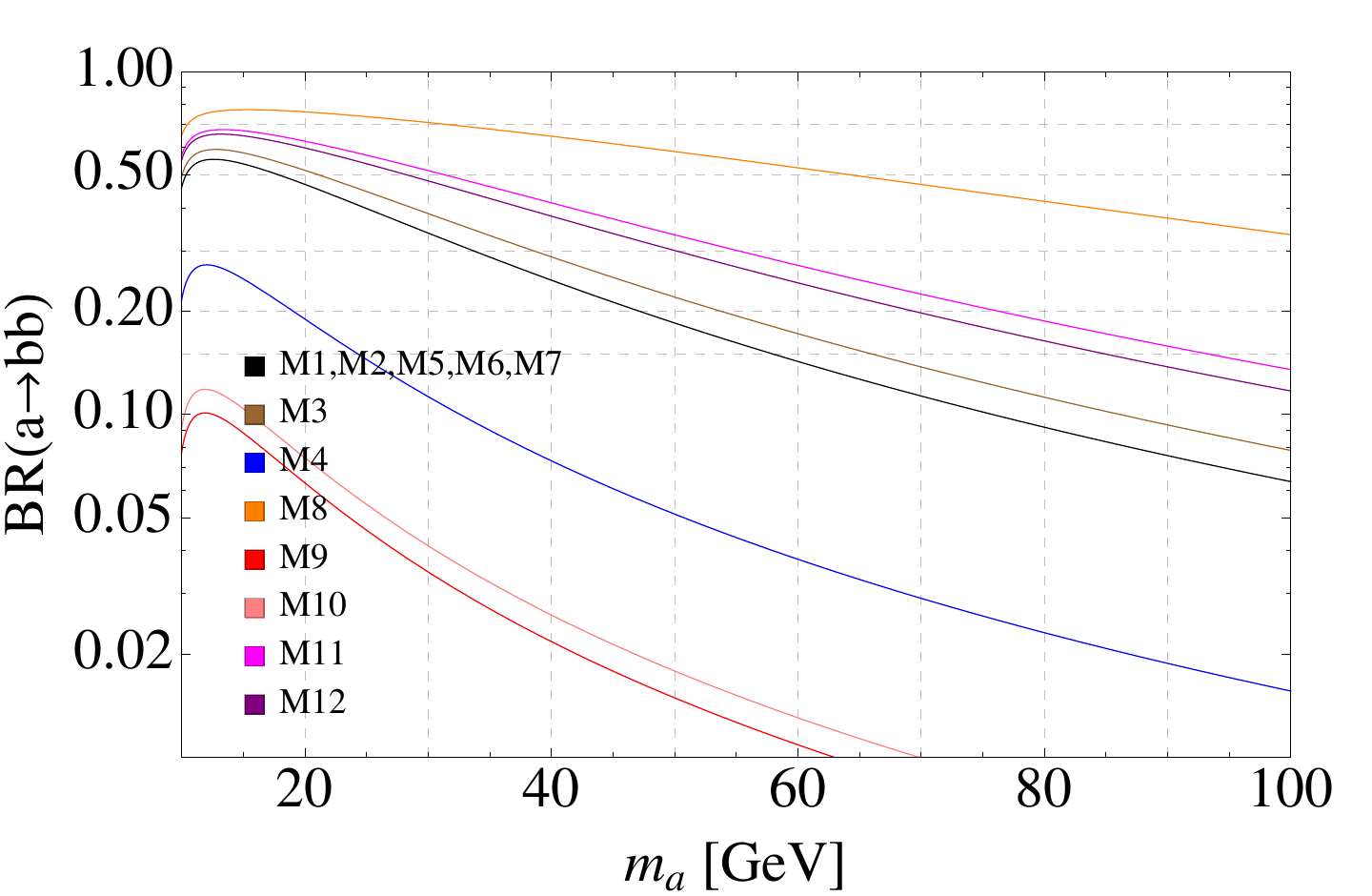}&
\includegraphics[width=0.5\textwidth]{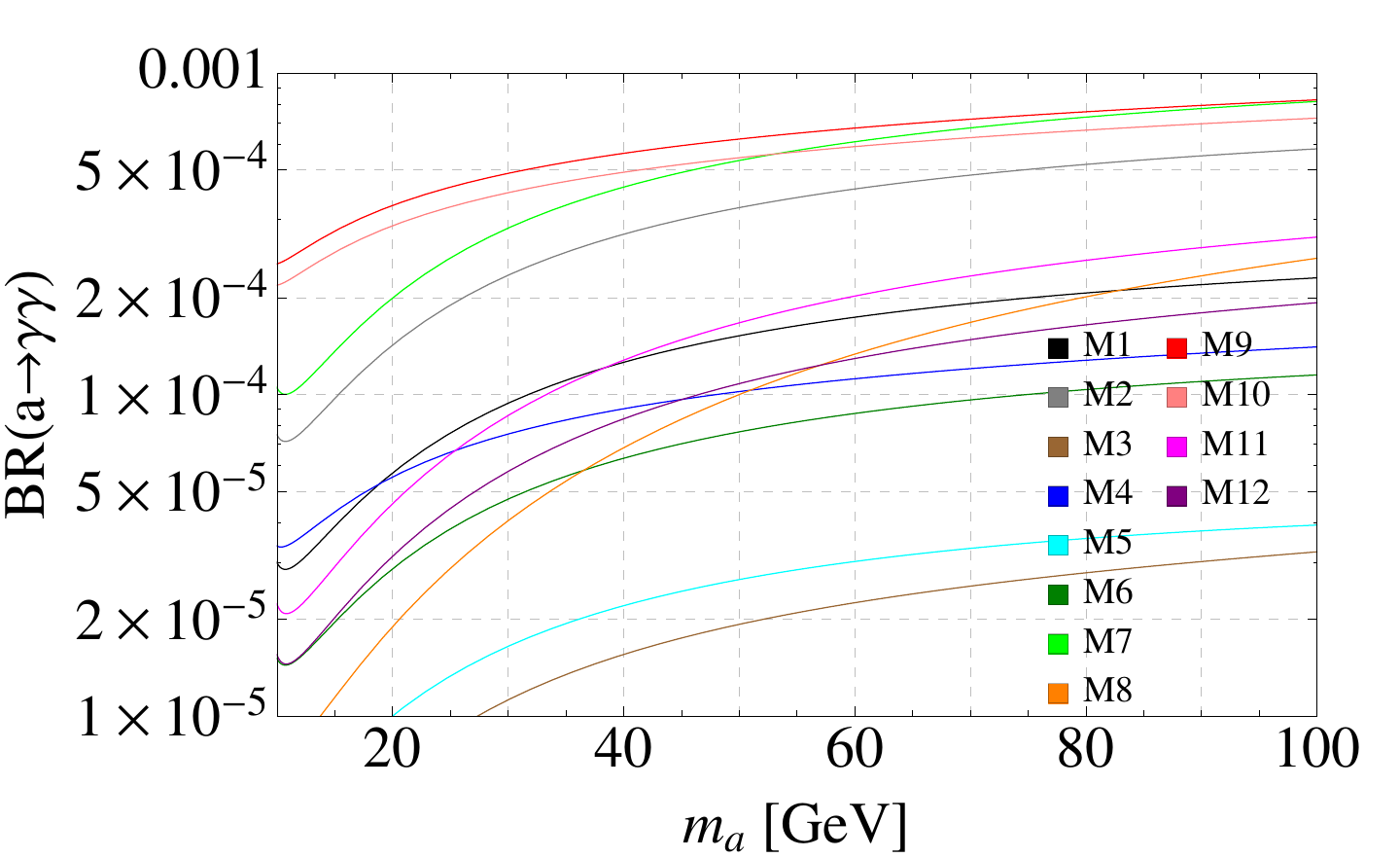}
\end{tabular}
\caption{Main Branching Ratios of the TCP for the reference models M1-M12 (c.f. Table~\ref{allmodels}) as a function of $m_a$: $a \to \tau^+\tau^-$ (top-left), $a\rightarrow$ hadrons (top right), $a\rightarrow b\bar{b}$ (bottom-left) and $a\rightarrow  \gamma\gamma$ (bottom-right).}
\label{fig:BRs}
\end{figure*}

In Figure~\ref{fig:prod} we plot the TCP production cross section for gluon fusion at the LHC with a center-of-mass energy of 7, 8 and 13 TeV. We also show, in Figure~\ref{fig:BRs}, the branching ratios in the main decay channels.

\section{Details of the simulation}\label{app:simulation}

As already stated in the main text, we generate a sample of signal events $p p\to a \to \tau^+ \tau^-$ for $m_a=10, 20, \cdots 100$~GeV  with up to two jets at the partonic level using MadGraph/ MadEvent. We shower and hadronize with Pythia and pass the resulting events through the fast detector simulation of Delphes using the standard CMS card with the modification on electron and muon isolation to be discussed below.

For the signal sample we set a $p_T$-cut of 100~GeV on the first jet at partonic level in order to increase the efficiency of the analysis at the reconstructed level. We use MLM matching and {\tt nn23lo1} PDFs for both signal and background generation.

First, we generate the signal samples with\footnote{Here, $K_{g, \rm eff} \sim K_g - 1/2 C_t$ includes the effect of top loops in the gluon-fusion production.} $K_{g,\rm eff}=C_\tau=1$ and $f_a=1$~TeV with the same pNGB width of 1~GeV for all masses and re-scale the production cross-section multiplying by ${\mathrm{GeV}}/\Gamma_{\tau\tau}$ for each model.

We then perform the cuts discussed in the main paper and below on these samples.
This gives the value of the TCP production cross-section after the cuts for $K_{g,\rm eff}=1$ and $f_a=1\mbox{ TeV}$ that we denote by $\bar \sigma_{\mathrm{prod.}} \times \epsilon$, where $\epsilon$ is the efficiency of the cut flow.

The true value of $\sigma_{\mathrm{prod.}}\times {\mathrm{BR}}_{\tau\tau}\times  \epsilon$ displayed in Table~\ref{Tab:signal} for each model is obtained by multiplying  $\bar\sigma_{\mathrm{prod.}}\times  \epsilon$ by $K_{g,\rm eff}^2 \times {\mathrm{BR}}_{\tau\tau}$ shown in Table~\ref{Kg2BR}. We do not include a $k$-factor for this analysis.
The efficiencies of the cuts depend on $m_a$ but are independent on the type of model.
Thus the expected signal $S$ is obtained as
$S = \sigma_{\mathrm{prod.}}\times {\mathrm{BR}}_{\tau\tau} \times \epsilon \times L \times ({\mathrm{TeV}}^2/f_a^2)$, where $L$ is the total luminosity.

\begin{table*}[h]
\begin{center}\begin{tabular}{c|c|c|c|c|c|c|c|c|c|c|}
  \hline
$m_a$ [GeV]& 10 & 20  & 30  & 40  & 50 & 60 & 70 & 80 & 90 & 100  \\
  \hline\hline
 {M1} & 6.7 & 3.4 & 2.5 & 1.9 & 1.5 & 1.2 & 0.95 & 0.79 & 0.66 & 0.57 \\
 {M2} & 9.7 & 4.8 & 3.6 & 2.7 & 2.1 & 1.7 & 1.4 & 1.1 & 0.96 & 0.82 \\
 {M3} & 5.7 & 2.9 & 2.2 & 1.8 & 1.4 & 1.1 & 0.91 & 0.76 & 0.65 & 0.56 \\
 {M4} & 6.2 & 2.6 & 1.6 & 1.1 & 0.79 & 0.60 & 0.47 & 0.39 & 0.32 & 0.27 \\
 {M5} & 3.0 & 1.5 & 1.1 & 0.84 & 0.66 & 0.52 & 0.42 & 0.35 & 0.30 & 0.25 \\
 {M6} & 3.0 & 1.5 & 1.1 & 0.84 & 0.66 & 0.52 & 0.42 & 0.35 & 0.30 & 0.25 \\
 {M7} & 9.7 & 4.8 & 3.6 & 2.7 & 2.1 & 1.7 & 1.4 & 1.1 & 0.96 & 0.82 \\
 {M8} & 0.88 & 0.50 & 0.48 & 0.46 & 0.43 & 0.40 & 0.36 & 0.33 & 0.30 & 0.28 \\
 {M9} & 1.9 & 0.74 & 0.42 & 0.27 & 0.19 & 0.14 & 0.11 & 0.091 & 0.076 & 0.064 \\
 {M10} & 1.8 & 0.73 & 0.41 & 0.27 & 0.19 & 0.14 & 0.11 & 0.091 & 0.076 & 0.064 \\
 {M11} & 2.1 & 1.1 & 0.94 & 0.79 & 0.66 & 0.55 & 0.47 & 0.40 & 0.35 & 0.30 \\
 {M12} & 2.9 & 1.5 & 1.3 & 1.0 & 0.85 & 0.70 & 0.59 & 0.50 & 0.43 & 0.37 \\
  \hline
\end{tabular}
\end{center}
\caption{Values of $K_{g,\rm eff}^2 \times BR_{\tau\tau}$ for the models M1 - M12 and $m_a=10\cdots 100$~GeV. For the values given, we chose the discrete charges of the top partners to be the same as those of the other fermions.}
\label{Kg2BR}
\end{table*}

\begin{table*}
	\begin{center}\begin{tabular}{|c||c|c|c|c|c|c|c|c|c|c|}
			\hline
			$m_a$ & 10 & 20 & 30 & 40 & 50 & 60 & 70 & 80 & 90
			& 100 \\
			\hline\hline
			\text{M1} & 9.9 & 12. & 9.0 & 6.5 & 5.3 & 3.7 & 3.0 & 2.3 & 1.7 & 1.4 \\
			\text{M2} & 14. & 17. & 13. & 9.3 & 7.7 & 5.4 & 4.4 & 3.2 & 2.4 & 2.0 \\
			\text{M3} & 8.5 & 10. & 8.2 & 6.0 & 5.0 & 3.6 & 2.9 & 2.2 & 1.6 & 1.4 \\
			\text{M4} & 9.1 & 9.2 & 5.9 & 3.8 & 2.9 & 1.9 & 1.5 & 1.1 & 0.80 & 0.67 \\
			\text{M5} & 4.5 & 5.2 & 4.0 & 2.9 & 2.4 & 1.7 & 1.4 & 1.0 & 0.74 & 0.63 \\
			\text{M6} & 4.5 & 5.2 & 4.0 & 2.9 & 2.4 & 1.7 & 1.4 & 1.0 & 0.74 & 0.63 \\
			\text{M7} & 14. & 17. & 13. & 9.3 & 7.7 & 5.4 & 4.4 & 3.2 & 2.4 & 2.0 \\
			\text{M8} & 1.3 & 1.7 & 1.8 & 1.6 & 1.6 & 1.3 & 1.2 & 0.96 & 0.76 & 0.69 \\
			\text{M9} & 2.7 & 2.6 & 1.5 & 0.93 & 0.70 & 0.46 & 0.36 & 0.26 & 0.19 & 0.16 \\
			\text{M10} & 2.7 & 2.5 & 1.5 & 0.93 & 0.70 & 0.46 & 0.36 & 0.26 & 0.19 & 0.16 \\
			\text{M11} & 3.1 & 3.9 & 3.4 & 2.7 & 2.4 & 1.8 & 1.5 & 1.2 & 0.87 & 0.74 \\
			\text{M12} & 4.3 & 5.3 & 4.6 & 3.5 & 3.1 & 2.3 & 1.9 & 1.4 & 1.1 & 0.92 \\
			\hline
	\end{tabular}\end{center}
	\caption{The values of $\sigma_{\mathrm{prod.}} \times BR_{\tau\tau}\times \epsilon$ in fb for $f_a = 1$~TeV for each of the models defined in Table~\ref{allmodels} if a seperation cut $\Delta R_{e\mu} > 0.1$ is imposed. The  main backgrounds are di-top and single-top
		(59.2~fb), $Z/\gamma^*$ (24.0~fb), and di-bosons (10.9~fb).}
	\label{Tab:signal01}
\end{table*}

\begin{table*}
	\begin{center}\begin{tabular}{|c||c|c|c|c|c|c|c|c|c|c|}
			\hline
			$m_a$ & 10 & 20 & 30 & 40 & 50 & 60 & 70 & 80 & 90
			& 100 \\
			\hline\hline
			\text{M1} & 0.86 & 4.7 & 6.8 & 5.6 & 4.9 & 3.5 & 2.9 & 2.2 & 1.7 & 1.4 \\
			\text{M2} & 1.2 & 6.8 & 9.8 & 8.1 & 7.1 & 5.0 & 4.2 & 3.2 & 2.4 & 2.0 \\
			\text{M3} & 0.74 & 4.1 & 6.1 & 5.2 & 4.7 & 3.3 & 2.8 & 2.1 & 1.6 & 1.4 \\
			\text{M4} & 0.79 & 3.7 & 4.4 & 3.3 & 2.7 & 1.8 & 1.5 & 1.1 & 0.80 & 0.66 \\
			\text{M5} & 0.39 & 2.1 & 3.0 & 2.5 & 2.2 & 1.6 & 1.3 & 0.99 & 0.74 & 0.62 \\
			\text{M6} & 0.39 & 2.1 & 3.0 & 2.5 & 2.2 & 1.6 & 1.3 & 0.99 & 0.74 & 0.62 \\
			\text{M7} & 1.2 & 6.8 & 9.8 & 8.1 & 7.1 & 5.0 & 4.2 & 3.2 & 2.4 & 2.0 \\
			\text{M8} & 0.11 & 0.71 & 1.3 & 1.4 & 1.4 & 1.2 & 1.1 & 0.94 & 0.76 & 0.68 \\
			\text{M9} & 0.24 & 1.0 & 1.1 & 0.81 & 0.65 & 0.44 & 0.35 & 0.26 & 0.19 & 0.16 \\
			\text{M10} & 0.23 & 1.0 & 1.1 & 0.81 & 0.65 & 0.43 & 0.35 & 0.26 & 0.19 & 0.16 \\
			\text{M11} & 0.27 & 1.6 & 2.6 & 2.4 & 2.2 & 1.7 & 1.5 & 1.1 & 0.86 & 0.74 \\
			\text{M12} & 0.37 & 2.2 & 3.5 & 3.1 & 2.9 & 2.1 & 1.8 & 1.4 & 1.1 & 0.92 \\
			\hline
	\end{tabular}\end{center}
	\caption{The values of $\sigma_{\mathrm{prod.}} \times BR_{\tau\tau}\times \epsilon$ in fb for $f_a = 1$~TeV for each of the models defined in Table~\ref{allmodels} if a seperation cut $\Delta R_{e\mu} > 0.2$ is imposed. The  main backgrounds are di-top and single-top
		(55.9~fb), $Z/\gamma^*$ (22.6~fb), and di-bosons (10.5~fb).}
	\label{Tab:signal02}
\end{table*}

The $Z/\gamma^*\to \tau \tau$ background cross-section is estimated from the Monte Carlo. We generate this sample in exactly the same way as the signal sample, i.e. with up to two jets MLM matched and a $p_T$ cut of 100~GeV on the leading jet at partonic level. We find the cross-section after matching to be 49~pb.

For the remaining background processes we generate  all the fully leptonic channels. We use the total cross-sections published by ATLAS, multiplied by the appropriate leptonic branching ratios for the $W$ and $Z$ (${\mathrm{BR}}(W\to l \nu) = 0.326$, ${\mathrm{BR}}(Z\to l^+ l^-) = 0.101$, where $l=e,\mu,\tau$, assuming ${\mathrm{BR}}(t \to W b)=100\%$. We find
$\sigma_{tt,{\mathrm{lep.}}} = 82.9~{\mathrm{pb}}$,
$\sigma_{tW,{\mathrm{lep.}}} = 10.~{\mathrm{pb}}$,
$\sigma_{WW,{\mathrm{lep.}}} = 15.1~{\mathrm{pb}}$,
$\sigma_{WZ,{\mathrm{lep.}}} = 1.66~{\mathrm{pb}}$,
$\sigma_{ZZ,{\mathrm{lep.}}} = 0.175~{\mathrm{pb}}$.

For the $(e,\mu)$ channel the QCD background and the single vector boson production+fakes background are expected to be sub-leading with respect to the irreducible backgrounds above. These can only be reliably computed by the experiment using data driven techniques. An order of magnitude estimate using a fake-rate of $10^{-3}$ for $j\to e$ and $10^{-4}$ for $j\to \mu$ and (conservatively) efficiencies similar to those of the signal sample $\approx 0.003$ leads us to estimate their total contribution after cuts to be at most a few fb.
The $t \bar t$ and $W t$  backgrounds are further reduced by imposing a $b$-jet veto, while the di-boson backgrounds containing a $Z$ boson are further reduced by vetoing on a third lepton. Neither of these last two cuts has any significant effect on the signal sample.

We now would like to pick a set of cuts that maximizes the figure of merit $\mathcal{Z}$ defined in Eq.~(\ref{sigmafom}) for our channel (opposite sign, opposite flavor di-lepton channel). We need to satisfy both trigger and isolation requirements for the leptons. From the trigger menus discussed in \cite{ATLAStrigger} we chose to retain only muons with $p_{T\mu} > 52$~GeV (off-line selection). This allows us to go rather low in the selection of $p_{Tj}$ of the leading jet, on which we are not triggering. We find that $p_{Tj}>150$~GeV gives an acceptable compromise between the signal rate an the reach in $\mathcal{Z}$. The $p_{Te}$ of the electron is chosen to be above 10~GeV and we use the standard isolation requirements between leptons and jets  $\Delta R_{l j}>0.5$. Furthermore, we put a third lepton veto and a $b$-jet veto in order to reduce the top and di-boson background.

The separation between the electron and the muon requires a more detailed discussion. Figure~\ref{dR} shows the $\Delta R_{e\mu}$ distribution of the signal (for two reference masses) and the main backgrounds before imposing $\Delta R_{e\mu}$ cuts. The background distributions of di-tops and di-bosons are rather flat while the signal -- especially for light TCPs --  is characterized by small $\Delta R_{e\mu}$. Thus, the sensitivity of the search is increased by imposing an \emph{upper} cut which we choose at  $\Delta R_{e\mu} < 1.0$.
 
In current searches, the ATLAS and CMS collaborations typically also use a \emph{lower} cut of  $\Delta R_{e\mu} > 0.1$ or $0.2$  as lepton isolation requirement. For these values, efficiencies and systematic errors are known and determined from data driven methods. Imposing an analogous isolation cut thus gives a reliable estimate of systematics and the reach of the study, but it comes at a price because for a light TCP, the signal efficiency is reduced. To study the impact of lepton isolation in more detail we explore three cases: without $\mu$-$e$ isolation, with $\Delta R_{e\mu}>0.1$, and $0.2$.  The resulting values for $\sigma_{\mathrm{prod.}}\times {\mathrm{BR}}_{\tau\tau}\times  \epsilon$ for the case of no isolation are given in Table~\ref{Tab:signal}. The analogous values with an isolation criterion  $\Delta R_{e\mu} > 0.1$ and $0.2$ are shown in Tables~\ref{Tab:signal01} and \ref{Tab:signal02}, respectively. 

The  background cross sections are barely affected by the modified isolation cuts. They are given in the caption of the respective table. The signal cross sections for the different isolations are comparable for a high mass TCP, but get reduced for $m_a\lesssim 20$~GeV (30~GeV) for $\Delta R_{e\mu} > 0.1$  $(0.2)$. Thus removing or at least reducing the isolation cut is advantageous for the TCP low mass regime if systematic errors can be kept under control.

Additionally, one could consider other channels such as  ($\tau_h,\mu$) and ($\tau_h,e$). Perhaps even the ($\mu,\mu$) channel could be relevant in spite of the large $Z / \gamma^*$ background. We refrain from doing this since the systematic error and the backgrounds become harder to estimate with our tools.

\section{Complementarity of $\tau\tau$ and $\gamma\gamma$ searches}\label{app:compare}

\begin{figure*}[tbh]
	\includegraphics[width=\textwidth]{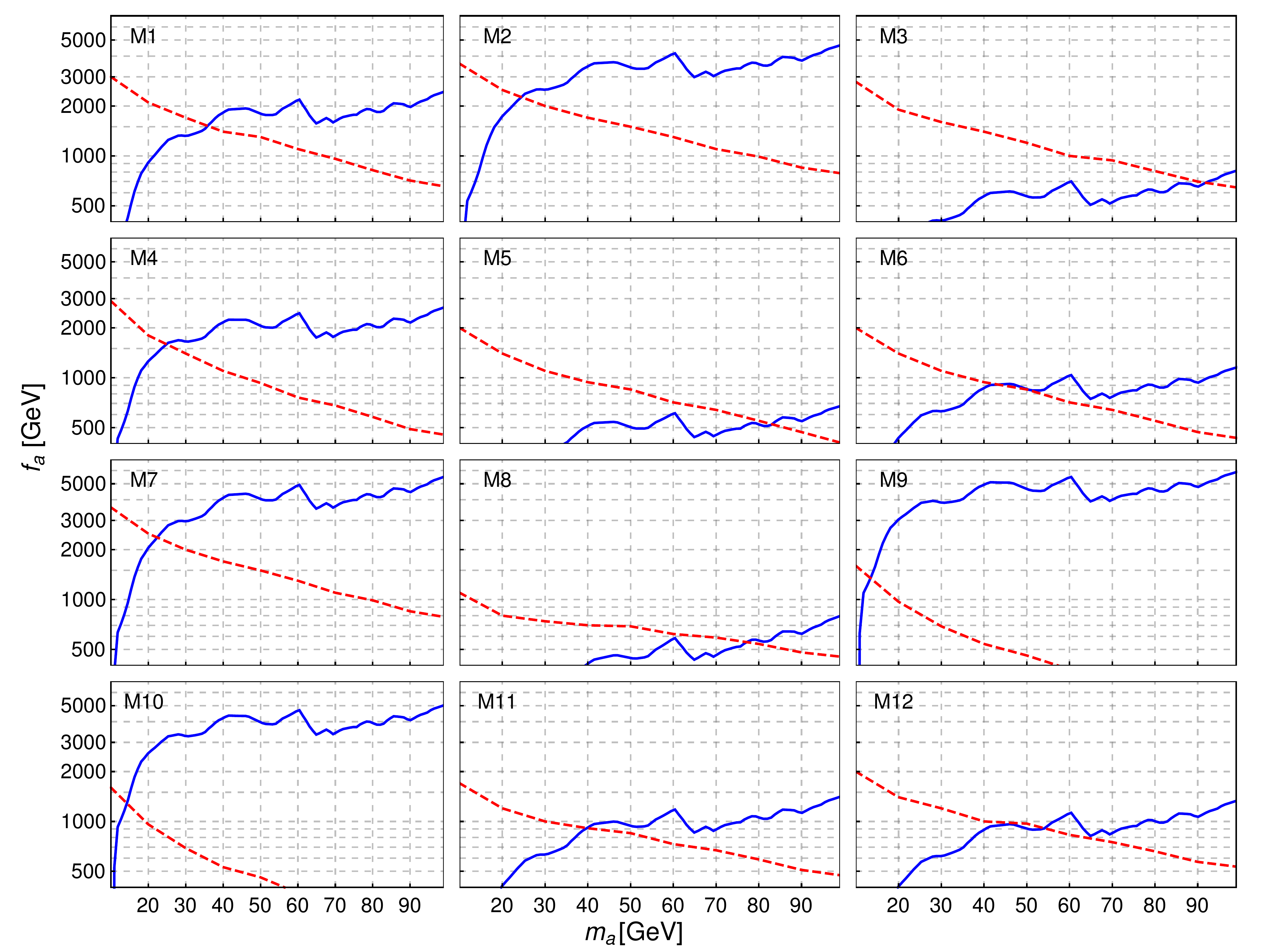}
	\caption{Projected bounds on $f_a$ from a boosted di-tau search with $\sqrt{s} = 13$~TeV after an integrated Luminosity of 300 fb$^{-1}$ (red dashed, from Fig. \ref{ditau}) and from a boosted di-photon search with $\sqrt{s} = 14$~TeV after an integrated Luminosity of 300 fb$^{-1}$ (blue solid, projections following Ref.~\cite{Mariotti:2017vtv}). }
	\label{fig:compare}
\end{figure*}

To conclude, we show in Fig.~\ref{fig:compare} a comparison of the reach of the di-photon analysis proposed in \cite{Mariotti:2017vtv} to the di-tau search proposed in this paper. We see that there is a nice complementarity between the two approaches, the di-tau being more sensitive in the low mass region and the di-photon nicely covering the high mass regions left uncovered. The complementarity also extends to the models themselves, namely some models, like M5 and M8, are much more sensitive to the di-tau signal, while others like M9 and M10 are covered by the di-photon analysis. A combination of the two approaches would essentially allow to test all of these models in the full mass range.

\bibliographystyle{JHEP}
\bibliography{lightPNGB}

\end{document}